\documentstyle[11pt,newpasp,twoside,epsf]{article}
\def\edcomment#1{\iffalse\marginpar{\raggedright\sl#1\/}\else\relax\fi}
\reversemarginpar

\begin{document}
\title{Sub--parsec to Mega--parsec jet emission and power} 
\author{Gabriele Ghisellini} \affil{Osservatorio Astron. di Brera,
V. Bianchi 46, I-23807 Merate, Italy }
\author{Annalisa Celotti} \affil{S.I.S.S.A., via Beirut 2-4, I-34014
Trieste, Italy}

\begin{abstract}
Relativistic jets carry a significant fraction of the total energy budget of 
a radio source, rivaling the power that is extracted through accretion.
A minor part of this bulk kinetic power is transformed to radiation, 
possibly through internal shocks if the plasma is accelerated, at the base 
of the jet, to a velocity which changes in time.
In this way we can understand why some radiation is produced all along the 
jet even if most of it originates at a preferred location, and why the 
efficiency of conversion of bulk to random energy is small.
The recent observations by Chandra of intense jet X--ray emission at large 
scales suggest that at least the ``spine" of jets continues to be highly 
relativistics even up to hundreds of kiloparsecs away from the nucleus
and give tight lower limits on the jet bulk kinetic power.
\end{abstract}

\section{Introduction}

The formation, acceleration and collimation of extragalactic
jets is still an open issue, despite several years of
active research, and despite the great amount of information
provided by improved VLBI techniques from the ground and now from
space, and the detection of jets in the optical and X--ray bands.
We still do not know the amount of power that the
jets must carry and their matter content.

Radio lobes, acting as calorimeters, should be the best site 
to measure the average energy supply, but even in this case the 
estimates depend on the uncertain assumption of equipartition 
and on the unknown amount of the proton contribution to the total
energy and pressure (Rawlings \& Saunders 1991).

The alternative is to measure the kinetic jet power $L_{\rm jet}$ 
directly, taking advantage of the apparent superluminal speeds measured 
in blazars to estimate their bulk Lorentz factor, and the minimum 
amount of leptons required to account for the observed emission.
This has been done by Celotti \& Fabian (1993) using radio 
data on the milli--arsec scale (corresponding to a linear 
scale of the order of a parsec).
Recently, the discovery that blazars are strong $\gamma$--ray 
emitters provided a new tool to measure $L_{\rm jet}$ on the 
sub--pc scale, and, even more recently, the discovery of relatively 
strong large scale X--ray jets by Chandra allowed yet another method 
to estimate $L_{\rm jet}$, this time on the 100 kpc -- Mpc scale.
The last two methods are based on spectral modeling of the 
spectral energy distribution (SED) of blazars, giving
the relevant estimates on the size, magnetic field,
particle density, and bulk Lorentz factor, necessary to 
calculate $L_{\rm jet}$, as detailed below.
For the implications regarding the jet formation
mechanism see Maraschi, these proceedings.

\section{Sub--pc scales}

Consider the properties of small scale jets, i.e. on a scale 
corresponding to the production of most of the blazar emission.
This scale has to be of order 10$^{-2}$ -- 10$^{-1}$ pc. 
It cannot be much smaller than this because of the required 
transparency of the source to the observed energetic $\gamma$--rays
(for the $\gamma\gamma \to e^\pm$ process)
and it cannot be much greater to account for the 
observed short variability timescales (Ghisellini \& Madau 1996). 
What can we infer on such scales?

Good spectral coverage of the broad band SED of a large number of 
blazars have allowed to characterize the
spectral properties of such objects: two broad peaks characterize
the blazar SED, and their peak frequencies and intensities
define different blazar subclasses (Padovani \& Giommi 1996).
Systematic trends in the SED have been then found (Fossati et al. 1998): 
both peak frequencies decrease for increasing bolometric power, and at 
the same time the high energy component becomes more dominant.

Recently, such a systematic behavior has been re-examined and
compared with the results obtained from the hard X--ray energy band of
a large number of blazars (Donato et al. 2001).  
In general a good agreement between the expected trend and the observed 
properties in the hard X--rays is found, although a quantitative
modification in the functional dependence of the spectral
characteristics on the power has been suggested (Donato et al. 2001).
In Fig. 1 we report the SED of blazars where data in each band
corresponds to the average luminosity of sources binned according to
their radio power (as derived by Fossati et al. 1998), with added the
new hard X--ray information and the analytical parametrization of the
SED proposed by Donato et al. (2001).

\begin{figure*}
\plotone{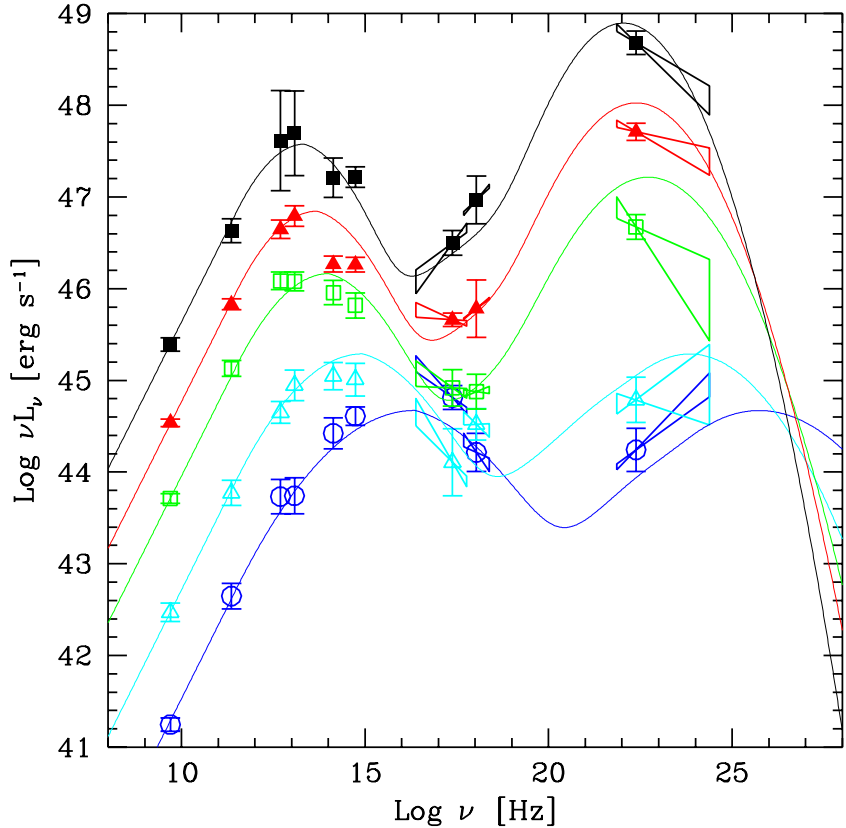} 
\vskip -0.5 true cm
\caption{Averaged SED of blazars. The data in each band are the result of
averaging over a number of sources, belonging to the 1 Jy complete sample
of BL Lacs, the Einstein SLEW survey sample of BL Lacs and the 2 
Jy complete sample of blazars, for a total of over 100 sources.
The sources have been divided into 5 bins of increasing 
radio power, thought to be representative of the bolometric one.
See Fossati et al. (1998) and the additional new collection of hard 
X--ray data in Donato et al. (2001).
The solid curves correspond to a phenomenological description of the 
average SED which is based on only one parameter (which can be the 
bolometric power, the peak synchrotron frequency or the radio power, 
which are related with one another through simple relations), as discussed
originally in Fossati et al. (1998) and slightly modified by
Donato et al. (2001).
}
\end{figure*}  

From the interpretational point of view, information on the high energy
component (most notably on the copious $\gamma$--ray emission) have
prompted the formulation of scenarios for the production of the blazar
broad band spectrum (e.g. Sikora, Begelman \& Rees 1994; Sikora 1994
and references therein). 
The phenomenological trends just discussed have been considered within 
such models. 
More specifically Ghisellini et al. (1998) have inferred physical properties 
of the sources from the modeling of a large number of blazar SED. 
The assumed scenario postulates that the emission is due to the synchrotron 
and inverse Compton processes (where the latter one acts on both the 
synchrotron photons themselves and any other externally produced radiation 
field) from a relativistically moving homogeneous source.

\subsection{The power of jets at the sub--pc scale}

Within such scenario, it is possible to estimate the relevant physical 
parameters of the emitting region, such as the size and  beaming 
factor, the magnetic field and the density of the emitting particles.
Therefore it is possible to calculate the flux of bulk kinetic and
magnetic energy transported by the jet.
It turns out that relativistic jets carry a significant fraction of the 
total energy budget of a radio source, rivaling the power that is 
extracted through accretion.  
In fact the bulk kinetic power of the emitting plasma largely 
exceeds the radiated one (see Fig. 2; Celotti \& Ghisellini, in prep).  
Such low efficiency of conversion of bulk to random (and then to radiative) 
energy is indeed expected if the dissipation is driven by the formation of
internal shocks, e.g. as those which would occur if the plasma is
accelerated, at the base of the jet, to a velocity which changes in
time (Ghisellini 1999; Spada et al. 2001). 
This plausible dissipation mechanism also naturally provides 
a preferred spatial location where the bulk of the radiation is 
produced (Rees 1978), similar to the one inferred from the arguments 
on source transparency to $\gamma$--rays and variability.

\begin{figure*}
\plotone{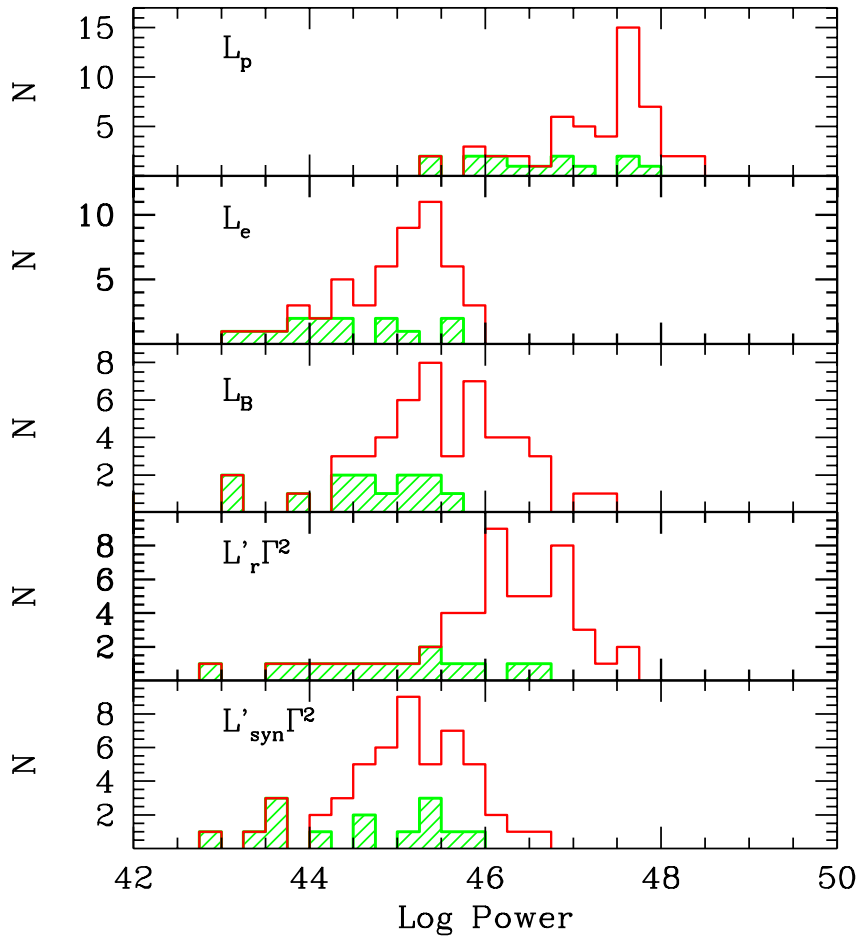} 
\caption{Histograms of powers estimated for flat spectrum radio loud
quasars and BL Lac objects (the latter ones represented by shaded areas.
Powers are in erg s$^{-1}$).  
$L^\prime_{\rm r}\Gamma^2$ and $L'_{\rm syn} \Gamma^2$ represent
the total and the synchrotron radiative power dissipated in the jet, 
respectively (where $\Gamma$ is the bulk Lorentz factor), while 
$L_e$, $L_p$ and $L_B$ indicate the kinetic powers associated with the 
electron component, the proton component (assuming one proton per 
electron), and the power transported as Poynting flux, respectively. 
The relative quantities are estimated by modeling the observed SED of 
blazars as due to synchrotron and inverse Compton emission from a 
homogeneous one--zone region (see Ghisellini et al. 1998 for details 
on the model). 
From Celotti \& Ghisellini, in prep.}
\end{figure*}  

\subsection{Electron--positron pairs?}

A further interesting point which can be inferred from the estimated
powers is the negligible role played by electron--positron pairs as
jet energy carriers. 
In fact, as shown in Fig. 2, the kinetic power associated with the 
relativistic emitting particles appears to be insufficient to provide 
the dissipated luminosity.  
A caveat should be discussed, namely the possible presence of particles 
emitting at energies below the observed frequency band, i.e. the 
extension/shape of the relativistic particle distribution to low 
energies, which can constitute a crucial uncertainty in the estimate 
of the bulk kinetic power (and thus the radiative efficiency) of jets
at the sub--pc scale.
As the particle distribution is typically steep, the particle number 
density -- and thus the bulk energy carried by the particles -- is 
crucially dependent on it (this is typically parametrized by the lower 
Lorentz factor of such energy distribution, $\gamma_{\rm min}$).

The good quality data in the soft X--ray band of powerful radio loud 
quasars provide tight limits on such quantity.  
If indeed -- as widely believed -- the X--to-$\gamma$--ray
component in such sources is dominated by the inverse Compton
scattering of externally produced photons (such as broad line or disk
photons) the soft X--ray spectrum is dominated by the photons
scattered by the lowest energy relativistic particles. 
The shape of the spectrum in this band allows then to determine the 
shape/extension of the lower end of the emitting particle distribution. 
As shown in Fig. 3 for one of the few blazars with data good enough to
accurately model the soft--medium X--ray emission, the soft spectrum
typically limits $\gamma_{\rm min}$ to be of order unity.

A second argument against a significant dynamical contribution of
electron--positron pairs in powerful radio--loud quasars derives from
the difficulty of producing them in sufficient number at the relevant
emitting jet scales. If pairs were produced in the inner compact
source, the surrounding intense photon field rapidly cools them,
enhancing the annihilation rate.  
The resulting surviving pairs, which can propagate along the jet, 
are numerically not enough to account for the required power (which 
has to exceed the radiated one) (Ghisellini et al. 1992; Celotti \& 
Ghisellini, in prep). 
Alternatively, pairs could be created along the jet and/or in the 
$\gamma$--ray emitting region itself.  
However, significant reprocessing of $\gamma$--rays into electrons 
and positrons would also lead to a copious emission from the pairs 
themselves in the X--ray band, well in excess of the observed X--ray 
flux (Ghisellini \& Madau 1996).

Note that these arguments imply that it is unlikely that the jet
plasma is dynamically dominated by pairs (i.e. that the kinetic
luminosity in pairs provides the bulk of the jet energy transport) but
do not exclude that (a smaller number of) pairs can contribute to the
emission. In particular the ratio of the proton power to the
radiatively dissipated one allows only $<$ 10 pairs per
electron/proton to be present. It is also worthwhile to stress that
the limits imposed by the presence of a high external radiation field
do not strictly apply to the weakest blazars (BL Lacs), for which
there is no strong direct evidence for the presence of a large density
of external photons.

\begin{figure*}
\plotone{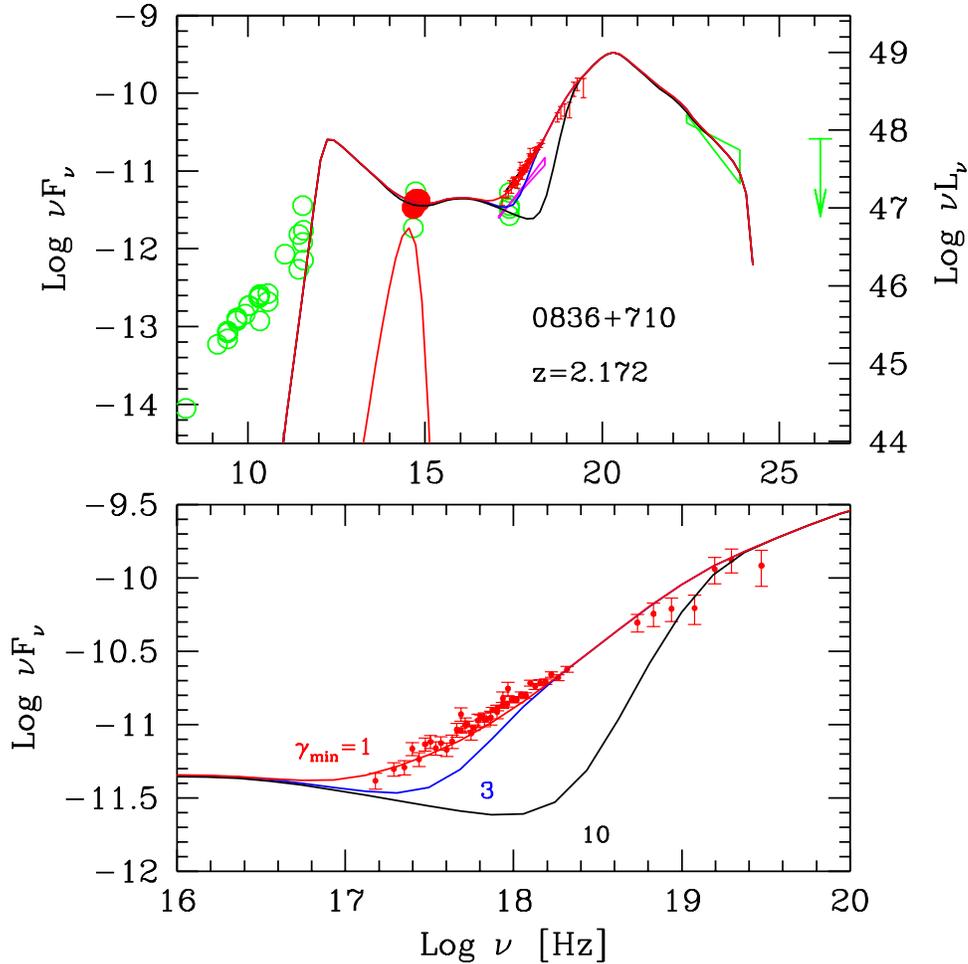}
\caption{The SED and a zoom on the X--ray spectrum ($Beppo$SAX data,
Tavecchio et al. 2000a) of the blazar 0836+710 
(upper and bottom panel, respectively).  
The lines represent the predictions from a model assuming that the two
spectral components are synchrotron and inverse Compton emission (both
synchrotron self--Compton and scattering of an externally produced
photon field, schematically represented as a peaked blackbody
component) from a homogeneous source (see Ghisellini et al. 1998 for
more details on the model). 
In particular in the bottom panel the model predictions are reported 
for different values of the lower Lorentz factor of the emitting particle 
distribution, $\gamma_{\rm min}$, which thus results well constrained by 
the soft X--ray data to values of order unity.
Note the hard X--ray emission dominates the power output (see the right
y--axis of the upper panel).}
\end{figure*}  


\begin{figure*}
\plotone{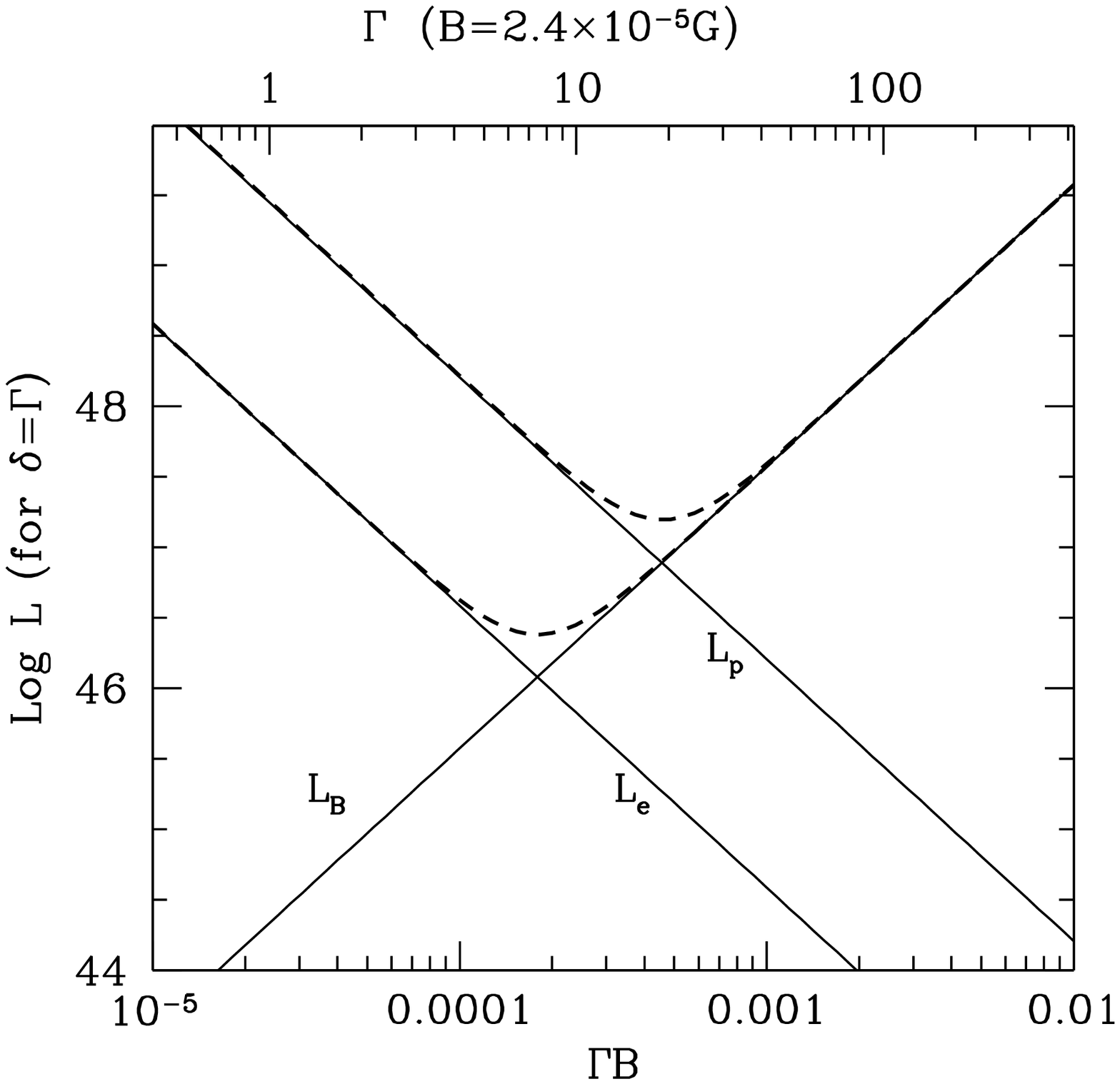} 
\vskip -0.5 true cm
\caption{Kinetic powers carried by protons ($L_{\rm p}$), emitting 
electrons ($L_{\rm e}$), and Poynting flux ($L_{\rm B}$), as
inferred from the large scale emission of the quasar PKS~0637--752 
as functions of the product of the bulk 
Lorentz factor and magnetic field ($\Gamma B$). 
The Doppler factor $\delta$ is assumed to be equal to the bulk Lorentz 
factor $\Gamma$, implying a viewing angle equal to $1/\Gamma$.
The electron number density carried by the jet is estimated
through the observed synchrotron luminosity. 
The kinetic powers $L_{\rm e,p} \propto (\Gamma B)^{-2}$,
while the Poynting flux $L_{\rm B} \propto (\Gamma B)^{2}$.
For $L_{\rm p}$ we assume one proton per electron.
The upper x--axis reports the values of $\Gamma$ assuming
the indicated magnetic field, resulting from fitting the
observed spectrum (Celotti, Ghisellini \& Chiaberge 2001). 
Note that the total transported power (sum of these components) is
minimized for bulk Lorentz factors similar to those
inferred on smaller scales, suggestively supporting the hypothesis
that at least part of the plasma flowing on large scales moves at
highly relativistic speeds. From Ghisellini \& Celotti (2001b).
}
\end{figure*}  

\subsection{What controls the blazar SED?}

A further piece of information recently emerged from the modeling of
the SED and the X--ray observations (by $Beppo$SAX) of the most
extreme, faintest, highly energy peaked BL Lacs.  
For the rest of the blazar population -- as for the sample considered 
in Ghisellini et al. (1998) -- a clear correlation was found 
between the energy of particles emitting at the energy peaks of the 
spectrum, $\gamma_{\rm peak}$, and the energy density $U$ (in magnetic plus
radiative fields), namely $\gamma_{\rm peak} \propto U^{-0.6}$.  
However, when extreme highly peaked BL Lacs (some of which detected in
the TeV energy band) are also considered, such correlation significantly 
steepens, becoming $\gamma_{\rm peak} \propto U^{-1}$
for small values of $U$.
The observed $\gamma_{\rm peak} \propto U^{-1}$ behavior,
for these blazars, is consistent with the internal shock scenario,
where the particle injection mechanism is not stationary,
but impulsive, and it lasts for a timescales comparable for
the time needed to one shell to cross the other.
During this time only the highest energy electrons radiatively cool,
steepening only the high energy part of the injected particle spectrum.
The rest of it retains its original slope.
In this case $\gamma_{\rm peak}$ does not correspond to the 
minimum energy of the injected electrons (as is the case
for more powerful sources), but to the energy for which 
$t_{\rm cool}(\gamma)=t_{\rm injection}$ (Ghisellini \& Celotti, in prep).

\section{Mpc--scales}

Radio, optical and X--ray observations have recently allowed huge 
progresses in the estimates and understanding of properties of  
large scale jets.
Most notably the detection by Chandra of intense X--ray emission at
100~kpc--Mpc scales has opened a new window to study the energetic and
physical processes occurring in jets and their interaction with the
environment.

In particular the X--ray data of the first detected source
(PKS~0637--752) and their comparison with information on similar scales
and at similar resolution in the radio and optical bands, support the
view that the X--rays are produced by inverse Compton scattering of
relativistic electrons against the cosmic microwave background
radiation (CMB).  
Alternative interpretations appear in fact to require more contrived 
conditions of the jet plasma (e.g. Schwartz et al. 2000; Celotti, 
Ghisellini \& Chiaberge 2001; Tavecchio et al. 2000b).  
The dominance of scattering on the CMB however requires that at least part 
of the emitting plasma is moving at highly relativistic speeds not only at
sub--pc and pc scales, but up to hundreds of kiloparsecs away from the
nucleus.  
This possibility appears to be at odds with radio observations implying 
at most only moderately relativistic velocities on the largest 
observed scales (such as the presence of both the jet and the counterjet).
But in fact the two sets of (X--ray and radio) findings might be easily 
reconciled and indicate the presence of a velocity structure in the jet, with 
a fast ``spine" surrounded by a slower ``layer'' -- i.e. a velocity gradient 
in the radial direction (see also e.g. Laing 1993, Chiaberge et al. 2000).

\subsection{The minimum power of large scale jets}

Further support to the hypothesis that jets are still moving
at highly relativistic speeds on the largest scales comes from
estimates of the transported powers. 
In fact -- at least for the best studied source so far, PKS~0637--752 -- 
the constraints on the plasma parameters inferred from the broad band 
distributions relative to hundreds of kiloparsecs scale emission, 
show that the total power (kinetic plus electromagnetic) associated 
with the emitting plasma is minimized -- for a given observed radiated 
luminosity -- for bulk Lorentz factors of order $\Gamma \sim 10$--20.  
The argument is simple: from the observed synchrotron power $L_{\rm s}$
we can estimate the (comoving) density of the emitting particles
$n^\prime \propto L_{\rm s}/[B^2 \delta^4]$.
The bulk kinetic power is therefore proportional to 
$L_{\rm e,p} \propto \Gamma^2 L_{\rm s}/[B^2 \delta^4]$, 
while the Poynting flux $L_{\rm B} \propto B^2\Gamma^2$.
Here $\delta$ is the Doppler factor, which is equal to the
bulk Lorentz factor for viewing angles close to $1/\Gamma$.
In this case $L_{\rm e,p}$ and $L_{\rm B}$ behave in a opposite way 
with respect to $\Gamma B$ and there is a minimum total power 
$L_{\rm e,p}+L_{\rm B}$ for some value of $\Gamma B$.
Fig. 4 reports the luminosity in the proton, electron and magnetic field 
components (and their sum as dashed lines) as a function of $\Gamma B$. 

Since the spectral fits yield an independent value of $B$, we can find
the value of $\Gamma$ which minimizes the jet power budget.
The found value of $\Gamma$ is fully consistent with those inferred 
from the spectral modeling and the jet speeds on nuclear scales 
(Ghisellini \& Celotti 2001a), as shown in the previous section.  

It should be finally stressed that -- according to this scenario --
information on the large scales provide tighter constraints with
respect to the sub--pc scales on the power estimates, as in the former
case the external radiation field intensity and spectrum (i.e. of the CMB)
can be robustly estimated.  

The presence of both a highly relativistic ``spine" and a slower 
layer in large scale jets implies that both blazars 
and radio--galaxies are expected to copiously radiate in the X--ray band 
through the inverse Compton process.  
In the case of radio--galaxies, in fact, the slow layer can be illuminated
by the boosted radiation coming from the nucleus, providing
extra seed photons for the inverse Compton process contributing
in the X--ray band (Celotti, Ghisellini \& Chiaberge 2001).
The emission from these slow layers is less beamed, and 
therefore visible also at large viewing angles (i.e. in radio--galaxies), 
while the strongly beamed emission from the spine is visible for 
aligned sources with a blazar--like core.

\section{Are jets more powerful than accretion disks?}

\begin{figure*}
\plotone{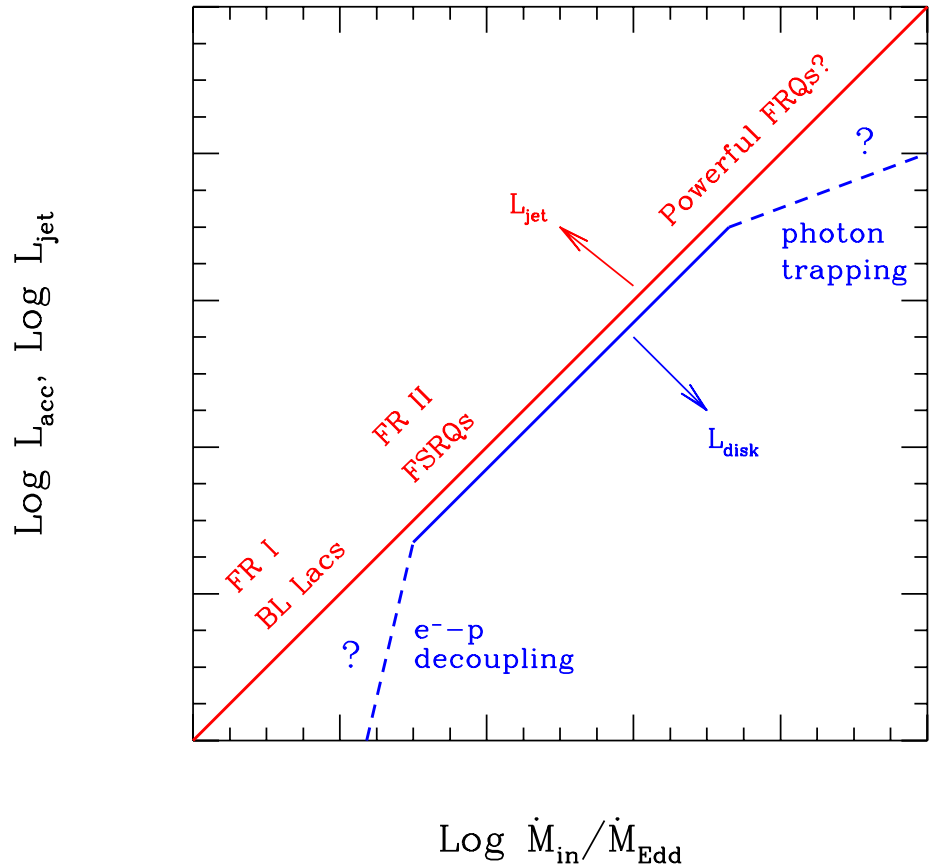} 
\vskip -0.5 true cm
\caption{Schematic diagram illustrating how the power of jets and
of accretion could scale as a function of $\dot m = 
\dot M_{\rm in} /\dot M_{\rm Edd}$.
Less powerful sources (BL Lacs and FR~I radio--galaxies) lacking broad 
emission lines should be characterized by radiatively inefficient 
accretion disks (protons do not transfer efficiently their energy to 
electrons) and in these sources the jet power may be dominant.
At the other extreme, at very large $\dot m$, the accretion disk could 
be again an inefficient radiator because of photon trapping,
and again the jet could more powerful than the disk.
At intermediate values of $\dot m$ the jet and accretion power are
more or less equal, as found by Rawlings \& Saunders (1991).
The fact that $L_{\rm jet} \propto \dot M_{\rm in}$ is justified by
the approximate equality between $L_{\rm jet}$ and $L_{\rm acc}$ for
the sources considered by Rawlings \& Saunders (1991), and by the fact
that the bulk Lorentz factor of all jets as estimated by their superluminal
velocities are distributed in a narrow range (see Jorstad et al. 2001).
}
\end{figure*}  

Rawlings \& Saunders (1991) suggested that $L_{\rm acc} \sim L_{\rm jet}$ 
for the considered FR~II and (a few) FR~I radio--galaxies
for which they could estimate
the (minimum) total energy in the radio lobes and their lifetime, yielding 
the average power supplied by the jet, i.e. $\langle L_{\rm jet}\rangle$,
and the luminosity in narrow emission lines, proportional to the ionizing
radiation coming from the disk and hence to $L_{\rm acc}$.
On the other hand there is little doubt that 
BL Lac objects (thought to be FR~I pointing at us)
are characterized by very weak or absent 
emission lines, invisible blue bumps, and relatively powerful jets.
For these objects therefore $L_{\rm jet}>L_{\rm acc}$.
At the other, high power, end of the sequence, there is again the 
indication that $L_{\rm jet}>L_{\rm acc}$, at least in a few cases,
such as PKS 0836+710.
We have collected these hints (admittedly not yet a robust scenario)
in a qualitative way in Fig. 5.

There are theoretical reasons to expect a deficit (with respect
to a pure linear proportionality) in the power extracted by
accretion and dissipated in the accretion disk at both power ends:
at high power photon trapping may prevent the produced radiation
to emerge from the accretion flow, and
at low power the $e$--$p$ decoupling can generate accretion disks
which are inefficient radiators, such as ion supported tori (Rees et 
al. 1982), advection dominated accretion flow
(ADAF, see e.g. Narayan, Garcia \& McClintock 1997), adiabatic
inflow--outflow (ADIOS, Blandford \& Begelman 1999) or a convection 
dominated flow (CDAF, Narayan, Igumenshchev \& Abramowicz 2000).
{\it Jets could therefore be the most efficient engines,}
and hints about their origin and acceleration may even come from
Gamma Ray Bursts (GRBs), whose radiation is probably collimated as well 
in a sort of jet (or ``flying pancake").
Their durations, in fact, indicate a relatively {\it long}
process ($10^4$--$10^5$ dynamical times), and it may be that the
same jet generation process is at work both in GRBs and
radio--loud AGNs.

\section{Conclusions}

We are still looking for the basic numbers of jets:
how much power they carry and what are they made of.
Progress has been made recently, and more is expected soon,
especially with high resolution observations in  radio, optical and 
X--rays of the same jet structures.

For instance, recent X--ray observations of radio galaxies embedded in 
clusters are start showing (in a few cases so far) a close connection 
between the morphology of the relativistic lobe components and the external 
thermal cluster gas. 
The high resolution images allow to improve the
estimates on the dynamical interaction of these two components which
in turn give significant constraints on both the jet matter content
and the filling factor of the relativistic plasma (Fabian et al. 2001). 
Complete disentangle of the values of these two quantities
can be foreseen with forthcoming deeper observations.

Another advance within immediate reach is the knowledge of the central 
black hole mass through velocity dispersion and/or optical luminosity 
of the host galaxy (Ferrarese \& Merrit 2000; Magorrian et al. 1998),
allowing to measure the jet powers in units of the Eddington luminosity.
This approach already allowed to interpret in a new way the division
line between FR~I and FR~II radio--galaxies in the radio--host optical
luminosity plane.
This can be due to a change in the accretion power as measured 
in units of the Eddington one: radio--galaxies above a critical 
value are FR~II, while FR~I are characterized, on average, by larger 
masses and lower accretion rates (Ghisellini \& Celotti 2001b).
Finally, and related to the difference between FR~I and FR~II
radio--galaxies, there might be important advances in 
numerical simulations, disclosing key features about shock
physics and about the problem of how jets (in FR~I sources)
are decelerated.

\acknowledgements The Italian MURST is acknowledged for financial
support (AC).

\end{document}